\documentclass[a4paper,11pt]{article}
\pdfoutput=1 
\usepackage{jheppub}

\newcommand{\CphiD}{C^{\varphi D}}
\newcommand{\QphiD}{Q_{\varphi D}} 
\newcommand{\Cphibox}{C^{\varphi \Box}}
\newcommand{\Qphibox}{Q_{\varphi \Box}} 
\newcommand{\Cphi}{C^{\varphi}}
\newcommand{\Qphi}{Q_{\varphi}} 
\newcommand{\Ceight}{C^8}
\newcommand{\phimin}{v}

\newcommand{\bea}{\begin{eqnarray}}
\newcommand{\eea}{\end{eqnarray}}

\usepackage{graphicx}
\usepackage{dcolumn}
\usepackage{bm}

\usepackage[
stretch=10, shrink=10, 
protrusion=true,
expansion=true
]{microtype}	

\DeclareMathOperator{\uu}{U}
\DeclareMathOperator{\su}{SU}

\newcommand{\spinor}[2]{\begin{pmatrix}#1 \\ #2\end{pmatrix}}

\newcommand{\dmu}{\partial_\mu}
\newcommand{\dmuup}{\partial^\mu}

\usepackage{xspace}
\newcommand{\GeV}{\ensuremath{\text{Ge\kern-0.15ex V}}\xspace}
\newcommand{\TeV}{\ensuremath{\text{Te\kern-0.1ex V}}\xspace}

\usepackage{csquotes} 
  \MakeOuterQuote{"} 
\usepackage{lmodern} 

\usepackage{xpatch}
\xpretocmd{\eqref}{Eq.~}{}{}

\usepackage{hyperref} 
\hypersetup{colorlinks=true,allcolors=black,linktocpage=true,breaklinks=true}

\title{A new perspective on the electroweak phase transition in the Standard Model Effective Field Theory}
\author{José Eliel Camargo-Molina,}
\emailAdd{eliel.camargo-molina@physics.uu.se}
\author{Rikard Enberg, and}
\emailAdd{rikard.enberg@physics.uu.se}
\author{Johan Löfgren}
 \emailAdd{johan.lofgren@physics.uu.se}
\affiliation{%
 Department of Physics and Astronomy, Uppsala University, Box 516, SE-751 20 Uppsala, Sweden
}%

\abstract{
A first-order Electroweak Phase Transition (EWPT) could explain the observed baryon-antibaryon asymmetry and its dynamics could yield a detectable gravitational wave signature, while the underlying physics would be within the reach of colliders. The Standard Model, however, predicts a crossover transition. We therefore study the EWPT in the Standard Model Effective Field Theory (SMEFT) including dimension-six operators. A first-order EWPT has previously been shown to be possible in the SMEFT. Phenomenology studies have focused on scenarios with a tree-level barrier between minima, which requires a negative Higgs quartic coupling and a new physics scale low enough to raise questions about the validity of the EFT approach.
In this work we stress that a first-order EWPT is also possible when the barrier between minima is generated radiatively, the quartic coupling is positive, the scale of new physics is higher, and there is good agreement with experimental bounds. Our calculation is done in a consistent, gauge-invariant way, and we carefully analyze the scaling of parameters necessary to generate a barrier in the potential. We perform a global fit in the relevant parameter space and explicitly find the points with a first-order transition that agree with experimental data. We also briefly discuss the prospects for probing the allowed parameter space using di-Higgs production in colliders.}

\begin{document}
\maketitle

\section{Introduction}
\noindent{}According to standard cosmological models, the electroweak symmetry of the Standard Model (SM) was broken in the early Universe through a phase transition from a phase of unbroken symmetry with a vanishing Higgs field, to a phase of broken symmetry with a non-zero expectation value. This phase transition is known as the electroweak phase transition (EWPT). Understanding the EWPT requires a detailed study of the vacuum of the theory, as described by the effective scalar potential at finite temperature. This is an important task for at least three reasons. 

The first reason is the matter-antimatter asymmetry of the Universe, which is not explained in the SM. One of the Sakharov conditions~\cite{Sakharov:1967dj} for dynamically generating a net baryon number is that there must be a departure from thermal equilibrium, and this could be provided by the EWPT if it is first-order. First-order phase transitions (FOPT) occur through bubble nucleation, which is a necessary condition to produce the asymmetry at the bubble walls. It must also be strongly first order, such that a generated asymmetry is not washed out by sphaleron processes. This proposed mechanism is known as electroweak baryogenesis~\cite{Kuzmin:1985mm,Morrissey:2012db}.

The second reason is that the bubbles nucleated during a first-order transition may, depending on the dynamics, generate a stochastic gravitational wave background (see Ref.~\cite{Weir:2017wfa} for a review) that has the potential to be detected by e.g.\ the LISA gravitational wave observatory~\cite{Caprini:2015zlo,Caprini:2019egz}.

Third, unlike other fundamental questions such as the nature of dark matter or dark energy (where there is no clear path towards a conclusive answer) a definitive answer on the order of the EPWT can be reached by setting bounds on deviations from the SM at the level that can be reached in future colliders \cite{Ramsey-Musolf:2019lsf}.

Given the significance of a possible first-order phase transition, it is somewhat disappointing that the SM does not allow one. The EWPT in the SM cannot be first-order if the Higgs mass is larger than about 70 GeV~\cite{Kajantie:1996mn}, because a barrier cannot be formed---a barrier to separate the minima is necessary for quantum tunneling, i.e., for first-order transitions. A first-order EWPT therefore requires some form of physics beyond the Standard Model (BSM) and is hence interesting from a phenomenological perspective:\ as emphasized in~\cite{Ramsey-Musolf:2019lsf}, particles with masses at or above the electroweak scale may affect the underlying physics of the EWPT. 

The EWPT has been studied extensively in two-Higgs doublet models (as first proposed in Ref.~\cite{Bochkarev:1990fx}, see e.g.\ \cite{Basler:2016obg} for more recent work). Other scenarios that have been studied are various scalar extensions (see e.g.\ \cite{Niemi:2018asa}) or models with strongly interacting sectors, e.g.\ composite Higgs models \cite{Bruggisser:2018mrt}. Such extensions often predict a strongly first-order EWPT.

Any modification of the scalar sector can in principle affect the possibilities for phase transitions. In addition, such modifications in general affect the triple Higgs self-coupling. There is intense ongoing activity in trying to probe this coupling, and thereby the whole scalar sector of the theory in di-Higgs production at the LHC (see \cite{DiMicco:2019ngk} for a thorough overview).

Instead of studying proposed models for physics beyond the SM one by one, it is useful to take a more model-independent approach. A way to do this systematically is to use effective field theory (EFT). The Lagrangian of the EFT contains a set of gauge-invariant, non-renormalizable operators, which are parametrized in powers of the inverse of a cutoff energy $\Lambda$. Because the operators are thought to be generated by physics at energies higher than the cutoff scale, the coefficients of these operators summarize deviations from the SM due to new physics at or above that scale in a consistent and gauge-invariant way.

In this paper we will consider the Standard Model Effective Field Theory (SMEFT)~\cite{Buchmuller:1985jz,Grzadkowski:2010es}, which to a given order in the expansion parameter $v/\Lambda$ contains all $\su(3)\times\su(2)\times\uu(1)$ invariant operators before electroweak symmetry breaking. We will work up to order $(v/\Lambda)^{2}$, i.e., we include operators up to dimension six. In the SMEFT, there are 59 such operators (for one fermion generation) that respect baryon number conservation. The Higgs sector of the SM is modified by three of these dimension-six operators, and these operators will be the main focus of our attention. 

These modifications to the Higgs sector also play a role in di-Higgs production at colliders. In this article we will study the EWPT in detail and we will discuss how it relates to di-Higgs production qualitatively, leaving a detailed study of the latter for future work.

The electroweak phase transition in models with dimension-six operators was first studied in \cite{Grojean:2004xa}, with later work in e.g.\ Refs.~\cite{Bodeker:2004ws,Huber:2007vva,Delaunay:2007wb,Cai:2017tmh,deVries:2017ncy,Chala:2018ari,Ellis:2018mja,Croon:2020cgk,Postma:2020toi}. The main focus in these studies is that a dimension-six operator in the scalar potential allows for a negative $\lambda$, which gives a tree-level barrier. This removes the need to generate a barrier radiatively, while still having a first-order phase transition. This mechanism requires a low cutoff, and recently it has been questioned whether such a low cutoff can be realized in extensions of the SM without invalidating the effective field theory framework~\cite{Postma:2020toi}.

While the case of a radiative barrier has been sometimes acknowledged in the literature \cite{Postma:2020toi,Ellis:2019flb, Croon:2020cgk}, one of the main results in the present article is that generating a barrier radiatively in the SMEFT is not only possible, but preferred from a phenomenology standpoint for several reasons. The new physics scale can be higher, the corresponding Wilson coefficients can be better probed in colliders and a very good agreement with the Higgs mass and electroweak precision tests can be very good.

In the SMEFT it is possible to obtain the observed Higgs mass of 125~\GeV with a much smaller (but still positive) value of $\lambda$ than in the SM. The value of $\lambda$ as compared to other couplings is the main determinant of whether it is possible to form a barrier in the potential, such that the phase transition becomes first-order \cite{Arnold:1992rz,Ekstedt:2020abj}. The reason that it is only possible to have a first-order phase transition in the SM for low Higgs mass is exactly this: a low Higgs mass in the SM translates to a small $\lambda$. But in the SMEFT we can achieve a small enough $\lambda$ because there are contributions to the Higgs mass from dimension-six operators.\footnote{An intriguing possibility is that $\lambda$ was small during the EWPT and then evolved to the SM value \protect\cite{Davoudiasl:2021vku}.}

We note that there are well-known problems with the calculation of the properties of the EWPT in perturbation theory. Two of those problems are that the effective potential is gauge dependent, and that the perturbative expansion breaks down close to the phase transition temperature. As described in more detail in section~\ref{sec:powercounting}, these issues can be solved simultaneously by using a consistent power counting~\cite{Ekstedt:2020abj}; in this paper we will use this method.

This paper is organized as follows.
In Section~\ref{sec:higgsSMEFT} we introduce the SMEFT and discuss the modifications introduced in the Higgs sector. In Section~\ref{sec:powercounting} we will describe the method of calculation and introduce the necessary power counting, and we will explain why there is no first-order phase transition in the SM. 
The possibility of first-order phase transitions in the SMEFT will be demonstrated in Section~\ref{sec:SMEFTPT} where we will also show that this is compatible with constraints on the SMEFT operators. In Section~\ref{sec:dihiggs} we discuss the prospects for probing this scenario in di-Higgs pair production at the LHC, while we leave a detailed study of this for the future. Finally we conclude in section~\ref{sec:conclusions}.

\section{The Higgs sector of the SMEFT}\label{sec:higgsSMEFT}
\noindent{}The SM has had striking scientific success in explaining and reproducing experimental results, but it is also a theory lacking answers to questions about a number of fundamental physical phenomena, such as the observed baryon asymmetry, the origin of neutrino masses or the nature of dark matter. This suggests that the SM is an excellent description of physics at the scales we have probed so far, but that new physics might live at much higher energy scales. In other words, there is a good chance that the SM is an effective field theory, the SMEFT. If that is the case, then the SM Lagrangian should include non-renormalizable interactions of dimension higher than four, suppressed by the scale of new physics but also contributing to observable phenomena.

There is only one dimension-five operator consistent with the SM gauge symmetries, the Weinberg operator $Q_5$. The story is much richer at dimension six. Up to equations of motion and total derivatives, it is possible to write down all possible operators, and thus write down an effective Lagrangian parametrizing all possible heavy new physics contributions as 
\begin{equation}
\mathcal{L}_{\text{SMEFT}} = \mathcal{L}_{\text{SM}} + C^{\text{5}} Q_{\text{5}} + \sum_{\text{dim 6}} C^i Q_i,
\end{equation}
where the sum is over all dimension-six operators $Q_i$. In this work we will use the Warsaw basis \cite{Grzadkowski:2010es}, and we will follow the notation of \cite{Dedes:2017zog}, where all Feynman rules are given in $R_\xi$ gauge for operators up to dimension six.

In particular, in \cite{Dedes:2017zog}, all Wilson coefficients are defined to have mass dimension $-2$, or in other words, the cutoff scale $\Lambda$ is not written explicitly. This is convenient as we do not have to assume any scale for new physics a priori. This also means that we do not take into account the effect of renormalization group running of the Wilson coefficients from $\Lambda$ to the EW scale\footnote{since our study does not require a large hierarchy between new physics and the electroweak scale, such effects are nevertheless sub-dominant.}. In the following we will work to order $(v/\Lambda)^{2}$. We refer to \cite{Dedes:2017zog} for a complete list of all SMEFT operators in the Warsaw basis. 

We want to understand whether a first-order EWPT is possible in the SMEFT. While all the dimension-six operators can in principle contribute to the calculations, the main actors will be those involving only Higgs and gauge fields. Doing so accounts for all new contributions to the classical potential, and for the wall at finite temperature, as we know from the SM calculation that the formation of a wall is driven by gauge boson loops~\cite{Arnold:1992rz,Ekstedt:2020abj}.

Thus, it turns out that it will be sufficient to focus on the operators involving only Higgs fields and covariant derivatives to show a first order transition is possible in the SMEFT. Operators including gauge boson fields explicitly are constrained to be small in comparison \cite{Ellis:2018gqa}, so in the following we focus on the operators 
\begin{align}\label{Eq:Ops}
\Qphi &= (\varphi^{\dagger}\varphi)^{3}, \\
\Qphibox &= (\varphi^{\dagger}\varphi)\square 
(\varphi^{\dagger}\varphi)\,, \\
\QphiD &= (\varphi^{\dagger} D_{\mu} \varphi)^{*} (\varphi^{\dagger} D^{\mu} \varphi).
\end{align}
and set the Wilson coefficients for any other operators to zero. Note that $\Qphibox$ is often written as $\dmu(\varphi^{\dagger}\varphi) \dmuup(\varphi^{\dagger}\varphi)$, which is related to the form used here through integration by parts. 
The Higgs Lagrangian is then given by
\bea\label{Eq:L}
\mathcal{L}_{\mathrm{H}} &= & (D_{\mu} \varphi)^{\dagger} (D^{\mu}
\varphi) + m^{2} (\varphi^{\dagger}\varphi) -
\frac{\lambda}{2}(\varphi^{\dagger}\varphi)^{2} +  \Cphi \Qphi \ +
\ \Cphibox \Qphibox \ + \CphiD \QphiD\; .
\eea
Here the Higgs doublet $\varphi$ is given by
\begin{equation}
\varphi=\spinor{\Phi^+}{\frac{1}{\sqrt{2}}\left(\phi+H+i\Phi^0\right)} \ ,
\end{equation}
where $\phi$ is a generic background-field and does not necessarily correspond to a minimum of the potential.\footnote{As we are using the conventions of Ref.~\protect\cite{Dedes:2017zog}, the quartic term in the Lagrangian (\protect\ref{Eq:L}) is multiplied by $\lambda/2$ instead of the more common $\lambda$, so care must be taken when interpreting our results. In our conventions, the SM value is $\lambda \approx 0.26$.}
The physical minimum $v$ of the classical potential satisfies the equation
\begin{equation}\label{eq:VEV-nonlinear}
-m^2+\frac{\lambda}{2}\phimin^2-\frac{3\Cphi}{4}\phimin^4=0.
\end{equation}
In principle there are several solutions though only one of them, 

\begin{equation}\label{V-nonlinear-solution}
 \phimin^2 = \frac{\lambda - \sqrt{\lambda^2 - 12 \Cphi m^2}}{3 \Cphi},
\end{equation}
corresponds to a nontrivial real-valued minimum. To linear order in $\Cphi$, the solution is
\begin{equation}\label{Eq:VEV}
    \phimin = {\phimin}_\text{SM}  + \frac{2 m^3}{\sqrt{2} \lambda^{5/2}} \Cphi \quad \mbox{ with }
    \quad {\phimin}_\text{SM} = \sqrt{\frac{2 m^2}{\lambda}}.
\end{equation}

The field $H$ above does not correspond to what we usually call the physical Higgs boson, because the derivatives in the operators $\Qphibox$ and $\QphiD$ in~\eqref{Eq:L} give a contribution to the kinetic term. To get canonically normalized kinetic terms, a field redefinition of the scalars gives a physical Higgs boson $h = Z_h H$ with 
\begin{equation}
Z_h^2 = 1 -2\Cphibox\phi^2 +  \frac{1}{2} \CphiD\phi^2.
\end{equation}
and a neutral Goldstone boson $G^0=Z_{G^0}\Phi^0$, with
\begin{equation}
Z_{G^0}^2 = 1 +  \frac{1}{2} \CphiD\phi^2.
\end{equation}
In order to compute the effective potential we need the masses of all particles for arbitrary values of the classical background field $\phi$. After linearizing in the Wilson coefficients, these are given by
\begin{align}
m_h^2   & = -m^2 + \frac{1}{2} \left(3 \lambda + m^2 (\CphiD - 4 \Cphibox) \right) \phi^2  - 
\frac{3}{4} \left(5 \Cphi + \lambda(\CphiD - 4 \Cphibox)\right) \phi^4 \label{eq:hmass}\\
m_{G^0}^2 & = -m^2 + \frac{1}{2} \left(\lambda+m^2\CphiD  \right) \phi^2
            -\frac{1}{4} (3 \Cphi+ \lambda\CphiD  ) \phi ^4 \label{eq:G0mass} \\
m_{G^+}^2 & = -m^2 + \frac{\lambda  \phi ^2}{2}-\frac{3 \Cphi \phi ^4}{4} \label{eq:Gpmass}\\
m_Z^2 & = \frac{(g^2 + {g'}^2) \phi^2}{4} \left(1 + \frac{1}{2}\CphiD \phi^2 \right) \label{eq:Zmass}\\
m_W^2 & = \frac{g^2 \phi^2}{4} \label{eq:Wmass}\\
m_t^2 &= \frac{y_t^2 \phi ^2}{2} \label{eq:tmass}.
\end{align}
We neglect all fermions except the top quark due to their very small Yukawa couplings. Note that, as mentioned in the beginning of this section, we have also set all Wilson coefficients except $\Cphi, \Cphibox$ and $\CphiD$ to zero, since we are here only concerned with the Higgs sector. 
In a complete study, the gauge boson masses additionally receive contributions from the $Q_{\varphi W}$, $Q_{\varphi B}$ and $Q_{\varphi WB}$ operators (see Ref.~\cite{Grzadkowski:2010es} for definitions) and the gauge couplings $g$ and $g'$ are rescaled by factors involving the first two of these. In addition, the top quark mass gets a contribution from the $Q_{u\phi}$ operator. 

The Higgs mass at the physical minimum is obtained by evaluating 
\begin{equation}\label{eq:hmass-nonlinear}
m_h^2=Z_h^{-2} \, \partial^2 V_0(\phi)|_{\phi=\phimin},
\end{equation}
which to leading order in the Wilson coefficients gives
\begin{align}
m^2_h &= 2m^2 \left[ 1 - \frac{2m^2}{\lambda^2} \left(3 \Cphi - 4\lambda \Cphibox + \lambda \CphiD\right)\right]\nonumber\\
&=  \lambda \phimin^2 - \left(3 \Cphi - 2\lambda \Cphibox + \frac{\lambda}{2} \CphiD\right) \phimin^4 .
\label{Eq:HiggsMass}
\end{align}
In the phenomenological analyses discussed in section~\ref{sec:SMEFTPT} we need the expressions for the Higgs mass and the location of the physical minimum. Note that the Goldstone bosons are massless at the minimum. It is also clear from \eqref{eq:Zmass} and \eqref{eq:Wmass} that the operator $\QphiD$ breaks custodial symmetry and its Wilson coefficient $\CphiD$ is therefore constrained to be relatively small. For that reason it is often omitted, but we will keep it in our analyses. In fact, as we will show in Section \ref{sec:SMEFTPT}, a non-zero (but small) $\CphiD$ can actually lead to a better fit to electroweak precision observables than in the SM. 

\section{Phase transition, gauge invariance, and power counting}\label{sec:powercounting}
\noindent{}To describe patterns of symmetry breaking in perturbative quantum field theory, one can study the scalar potential of the theory. To include quantum and thermal effects, one should use the effective potential, and typically this is done perturbatively. However, naive application of perturbation theory to the effective potential leads to problems such as gauge dependence and a breakdown of the perturbative expansion~\cite{Nielsen:1975fs,Patel:2011th,Arnold:1992rz,Andreassen:2014eha}. 

Thus, the usual method of calculating the critical temperature and strength of the phase transition, whereby the effective potential is minimized numerically, gives a gauge-dependent result. However, the critical temperature can be computed in a gauge invariant way by performing a consistent expansion in orders of $\hbar$~\cite{Patel:2011th}. But there is also the problem of perturbative breakdown for large temperatures. Curing this breakdown requires a resummation to all orders, which is not consistent with the perturbative expansion required for gauge invariance.\footnote{This problem is similar to that of Goldstone IR divergences of the zero temperature effective potential. See Refs.~\protect\cite{Elias-Miro:2014pca,Martin:2014bca,Espinosa:2017aew} for how to deal with these issues in Landau gauge.} In~\cite{Ekstedt:2020abj} a method for controlling both of these problems based on a careful power counting was developed, inspired by previous work in Refs.~\cite{Arnold:1992rz,Laine:1994zq,Patel:2011th}. In this paper we will use the method of Ref.~\cite{Ekstedt:2020abj} for our calculations.

The important insight of this approach is that the loop corrections to the tree level potential have to be large enough to change its shape. This means that the loop expansion is no longer applicable, and perturbation theory has to be reordered according to some other power counting scheme. We will here consider this, first for the SM, and then for the SMEFT.

\subsection{Power counting in the SM}\label{sec:powercountingSM}
\noindent{}The tree-level potential of the SM does not feature a barrier,
$$ V^{\textsc{SM}}_0(\phi) = -\frac{m^2}{2}\phi^2+\frac{\lambda}{8}\phi^4, $$
which means that a barrier would have to be induced by radiative corrections for a first-order phase transition to take place. The leading thermal corrections of a bosonic field with squared mass $x$, 
$$ 
f(x)=-\frac{\pi^2}{90}T^4 + \frac{T^2 x}{12} - \frac{T x^{3/2}}{12 \pi} + \mathcal{O}(T^0), 
$$
show that a barrier can be generated from the term $-T x^{3/2}/ 12\pi$. A bosonic particle with field-dependent mass squared $M^2\sim e^2 \phi^2$ would then give a contribution $\sim T e^3 \phi^3/ 12\pi$, which, being cubic in $\phi$, is precisely the ingredient we need so that a barrier can be generated.\footnote{Note that fermionic particles do not contribute such a term.}
The question is then whether these terms are large enough to generate a barrier, and to address this question we have to turn to power counting. With the leading thermal corrections included, the potential will be roughly of the form\footnote{Here, for simplicity, we neglect the contributions from the temporal gauge bosons \cite{Arnold:1992rz}. See \eqref{eq:VLOSMEFT} for the complete expression.}
\begin{equation}\label{eq:VLOSM}
   V^{}_{\textsc{LO}}(\phi,T) = -\frac{1}{2}m^2_{\text{eff}}(T) \phi^2 - e^3 \frac{T}{12 \pi}\phi^3 + \frac{\lambda}{8}\phi^4.  
\end{equation}
Here $m^2_{\text{eff}}(T)=m^2- \alpha T^2/12$, and $\alpha$ and $e^3$ are combinations of electroweak coupling constants, which in the SM are given by
\begin{align}
\alpha &= 3\lambda + \frac{9}{4} g^2 + \frac{3}{4} {g'}^2 + 3y_t^2, \\
e^3 &= \frac{1}{2} g^3 + \frac{1}{4} \left(g^2 + {g'}^2\right)^{3/2} \label{eq:e3}.
\end{align}
For a first-order phase transition to take place, all of the terms in \eqref{eq:VLOSM} should be equally important near the critical temperature. This balance has been worked out by Arnold and Espinosa~\cite{Arnold:1992rz}, and gives the following necessary power counting
\begin{equation}\label{Eq:scaling}
  \lambda \sim e^3:~\hspace{3em} \phi \sim T~\hspace{1em}\land~\hspace{1em}m^2_{\text{eff}}(T)\sim e^3 T^2~\hspace{1em}
  \land ~ \hspace{1em}T\sim \frac{\sqrt{m^2}}{e}.
\end{equation}

In the SM we know that the measured Higgs mass is too large to allow for a first-order phase transition~\cite{Kajantie:1996mn}. With the idea of power counting we can understand why: the size of the Higgs mass in the SM is tied to the size of $\lambda = m_h^2/\phimin^2$, and a large Higgs mass hence requires a large $\lambda$. But the existence of the barrier hinges on whether the power counting $\lambda \sim e^3$ applies or not. To understand the numerical values that are appropriate, we should compare the numbers in front of the $\phi^3$ and $\phi^4$ terms. Then we find that we want $\lambda \approx \frac{3e^3}{4 \pi}$ for the power counting $\lambda \sim e^3$ to apply and the barrier to manifest. Here $e$ is the coupling in \eqref{eq:e3}, or roughly $e \sim g$. By comparing the numbers it can be seen that a Higgs mass of $125~\GeV$ corresponds to a much too large $\lambda$ that more closely corresponds to $\lambda \sim e^2$, which forbids the existence of the barrier. 

There are hence three main ways to achieve a first-order phase transition in an extension of the SM. Either (1) the zero temperature potential has to be affected in a major way, e.g.\ with a tree-level barrier, or (2) there needs to be new bosonic fields 
with a stronger coupling to the Higgs background field than that of the Higgs itself, or (3) there has to be some mechanism that allows $\lambda$ to be smaller than in the SM.

\subsection{Power counting in the SMEFT}\label{sec:powercountingSMEFT}
\noindent{}The tree-level potential of the SMEFT contains a $\phi^6$ term that is not present in the SM,
\begin{equation}
    V_0(\phi)=-\frac{m^2}{2}\phi^2+\frac{\lambda}{8}\phi^4 - \frac{\Cphi}{8}\phi^6,
\end{equation}
which could conceivably change the nature of the phase transition. As has been noted before, it is possible to generate a tree-level barrier with a negative $\lambda$~\cite{Grojean:2004xa}. For a careful and gauge-invariant study of the tree-level barrier case in SMEFT, see~\cite{Croon:2020cgk}. Here we want to focus on a different possibility.  

As noted above, the SM can not simultaneously explain the measured value of the Higgs mass and have a first-order phase transition---$\lambda$ would be too large. In the SMEFT, however, the Higgs mass squared is not just proportional to the quartic coupling, but also has contributions from the new operators, see \eqref{Eq:HiggsMass}. However, the necessity of a small $\lambda$, with $\lambda\sim e^3$, is the same as in the SM. Any hope for a first-order phase transition with the right Higgs mass hence requires that the new contributions are at least comparable to the SM one. In other words, given the scaling in \eqref{Eq:scaling}, we want all terms in \eqref{Eq:HiggsMass} to be about the same order of magnitude: 
\bea
\left|\lambda \phimin^2 \right| \sim \left|3 \Cphi \phimin^4\right| \sim \left|2\lambda \Cphibox \phimin^4\right| \sim \left|\frac{\lambda}{2} \CphiD \phimin^4\right| ,
\eea
which implies
\begin{equation}
  \Cphi \sim \frac{\lambda}{\phimin^2}  \qquad \text{and} \qquad \Cphibox, \CphiD \sim \frac{1}{v^2}.
\end{equation}
Although it seems that the coefficients $\Cphibox, \CphiD$ have to be quite large (corresponding to a cutoff $\Lambda\sim \phimin$), it turns out that it is enough to have a sizable $\Cphi$ to reproduce the measured Higgs mass with a small but positive $\lambda$. The coefficients $\Cphibox, \CphiD$ can then be much smaller than this naive power counting suggests, and in fact, large values for these parameters are ruled out from experimental constraints, as described in Section~\ref{sec:SMEFTPT}. 

We can turn the estimation of $\Cphi$ into a natural value for the cutoff $\Lambda$,
\begin{equation}\label{eq:cutoff}
    \Cphi= \frac{1}{\Lambda^2} \sim \frac{\lambda}{\phimin^2} \implies \Lambda \sim \lambda^{-1/2} v.
\end{equation}
Interestingly, the relative sizes of these couplings means that linearizing in the dimension-six Wilson coefficients when calculating the physical Higgs mass and minimum does not yield a good approximation. This is not surprising because we are demanding that the terms in the potential be of equal size, and should not be negligible when compared to each other. For phenomenological studies we should hence use the non-linearized relations,~\eqref{V-nonlinear-solution} and~\eqref{eq:hmass-nonlinear}. This seems to suggest that operators of dimension eight and higher should also be included, although this turns out not to be the case. To demonstrate this, consider adding a dimension eight term $\Ceight (\varphi^\dagger\varphi)^4$ to the Lagrangian. The equation for the minimum,~\eqref{eq:VEV-nonlinear}, changes to
\begin{equation}
    -m^2+\frac{\lambda}{2}\phimin^2-\frac{3\Cphi}{4}\phimin^4-\frac{\Ceight}{2}\phimin^6=0.
\end{equation}
Inserting the natural expectation, $\Ceight \sim 1 / \Lambda^4$ with $\Lambda$ as in~\eqref{eq:cutoff}, shows that this term is subdominant with respect to the others---and should be neglected. The same argument holds for the Higgs mass.

Similarly, by comparing the sizes of the different terms near the critical temperature, we can estimate whether the dimension-six operator plays any other role during the phase transition. Then, if we note that $\phimin \sim \sqrt{m^2/\lambda}$ and $\phi \sim T \sim \sqrt{m^2}/e$, we have
\begin{equation}
    \lambda \phi^4 \sim e^3 \phi^4 \qquad \text{and} \qquad \Cphi \phi^6 \sim e^3 \phi^4 \frac{\phi^2}{\phimin^2} \sim e \times e^3 \phi^4.
\end{equation}
This shows that the $\Cphi \phi^6$-term is subleading in our power counting and is only relevant at higher orders in the perturbative expansion.

The other Wilson coefficients, including $\Cphibox$ and $\CphiD$, will enter the analysis through loops. Each such operator will be suppressed by a factor $\left(\phi/\phimin\right)^2 \sim e$ compared to the dimension-four operators, and will automatically be moved higher up in the perturbative expansion. This means that at leading order the effective potential is analogous to that of the SM, with minor modifications.

To show the cancellation of the gauge-dependent terms, we can consider $R_\xi$-gauge. Then the square mass of the Goldstone field and ghost corresponding to the $Z$ read
\begin{align}
m_{\chi}^2 & = m_{G^0}^2 + m_{\eta_Z}^2,\\
m_{\eta_Z}^2 & = \xi m_Z^2.
\end{align}
There are four modes which introduce gauge-dependence at one-loop order through their square masses: the Goldstone ($m_{\chi}^2$), the longitudinal gauge boson ($m_{\eta_Z}^2$), and the two ghost fields ($m_{\eta_Z}^2$). In the current power-counting, there is a separation of scales between $m_{G^0}$ and $m_Z$: $m_{G^0}^2/m_Z^2 \sim e$. This means that the possibly problematic contributions to the one-loop potential,
\begin{equation}
    -\frac{T}{12 \pi} \left((m_{\eta_Z}^2)^{3/2} + (m_{\chi}^2)^{3/2} - 2(m_{\eta_Z}^2)^{3/2} \right) = -\frac{T}{12 \pi}\left(\frac{3}{2} m_{G^0}^2 m_{\eta_Z} + \mathellipsis \right),
\end{equation}
are actually of higher order ($\sim e^4 T \phi^3$) than the leading order terms ($\sim e^3 T \phi^3$).

The full leading order potential is then gauge invariant, and reads
\begin{equation}\label{eq:VLOSMEFT}
   V^{}_{\textsc{LO}}(\phi,T) = -\frac{1}{2}m^2_{\text{eff}}(T) \phi^2 - e^3 \frac{T}{12 \pi}\phi^3 - \frac{T}{12 \pi}\left(2 W_L^{3/2}+Z_L^{3/2}+A_L^{3/2}\right) + \frac{\lambda}{8}\phi^4,
\end{equation}
where $W_L, Z_L, A_L$ are the resummed square masses of the temporal mode of the corresponding gauge boson (see e.g.\ appendix C of~\cite{Ekstedt:2020abj}), and $m^2_{\text{eff}}$ is given by
\begin{equation}
    m^2_{\text{eff}}=m^2-\left(3\lambda + \frac{9}{4} g^2 + \frac{3}{4} {g'}^2 + 3y_t^2+m^2 \left(\CphiD-2 \Cphibox\right)\right) \frac{T^2}{12}.
\end{equation}
We note that this is different from the tree-level barrier case (as e.g.~in \cite{Croon:2020cgk}) where the larger Wilson coefficients mean that more care has to be taken with resummations, with contributions proportional to $\Cphi T^4 \phi^2$ and  $\Cphi T^2 \phi^4$ appearing in the gauge-invariant leading order potential.

\section{First order phase transition in the SMEFT with $\lambda > 0$}\label{sec:SMEFTPT}
\noindent{}Using power counting we have shown that a first order phase transition with positive $\lambda$ is not only possible in the SMEFT but that it roughly corresponds to values of the Wilson coefficients that might be consistent with current experimental bounds.
However, a more solid statement requires a detailed study on both the order of the phase transition and the observable prediction for each parameter point in the relevant region of the SMEFT parameter space. 

By combining the rough estimate from power counting, constraints from the literature (see for example \cite{Ellis:2018gqa,Dawson:2020oco}), and a preliminary grid scan, we decided to focus on the parameter region (in units of $\GeV^{-2}$):
\bea \label{eq:sampling}
 \Cphi  &\in& [-1 \cdot 10^{-5} , 0]  \\ \nonumber
  \CphiD &\in& [-1 \cdot 10^{-7}, 0.5 \cdot 10^{-7}]  \\ \nonumber
\Cphibox &\in&   [-1.5 \cdot 10^{-6}, 3 \cdot 10^{-6}], 
\eea
while we also fix $\lambda$ and $m^2$ such that the Higgs mass $m_H$ and $\phimin$ agree with their measured values. 
We sampled the region using a $150\times 150 \times 150$ grid and kept only points with positive $\lambda$.

We then find the critical temperature $T_c$ and the minimum evaluated at the critical temperature $\phi_c\equiv\phi(T_c)$ numerically, using the leading order finite temperature potential in \eqref{eq:VLOSMEFT}. In this way we get the ratio $\phi_c/T_c$ which is commonly used to determine whether the phase transition is strongly first order, through the condition $\phi_c/T_c > 1$~\cite{Morrissey:2012db}. 

The numerical minimization of the leading order potential given in~\eqref{eq:VLOSMEFT} is performed with the implementation of the Nelder-Mead simplex algorithm in GSL~\cite{GSL}, for different temperatures. The critical temperature is then located through a binary search by first placing lower and upper bound on the temperature. The lower bound $T_0$ arises from the fact that the origin needs to be a minimum for the barrier to form, which is only true when $-m^2_{\text{eff}}>0$; with $T_0$ determined through $m^2_{\text{eff}}(T_0)=0$. The upper bound is simply a conservative estimate, $T_{\text{max}}=200~\GeV$, where the system is basically guaranteed to be in the symmetric phase. Then the interval $[T_0,T_{\text{max}}]$ is halved by testing which phase the midpoint $(T_{\text{max}}-T_0)/2$ is in. This determines the next step of the search; if this point corresponds to the symmetric phase then one halves the left interval, and vice versa. This procedure is repeated until the desired accuracy is obtained (within $10^{-6}$~\GeV in our case).

\begin{figure}[tbh!]%
    {{\includegraphics[width = 0.9\textwidth]{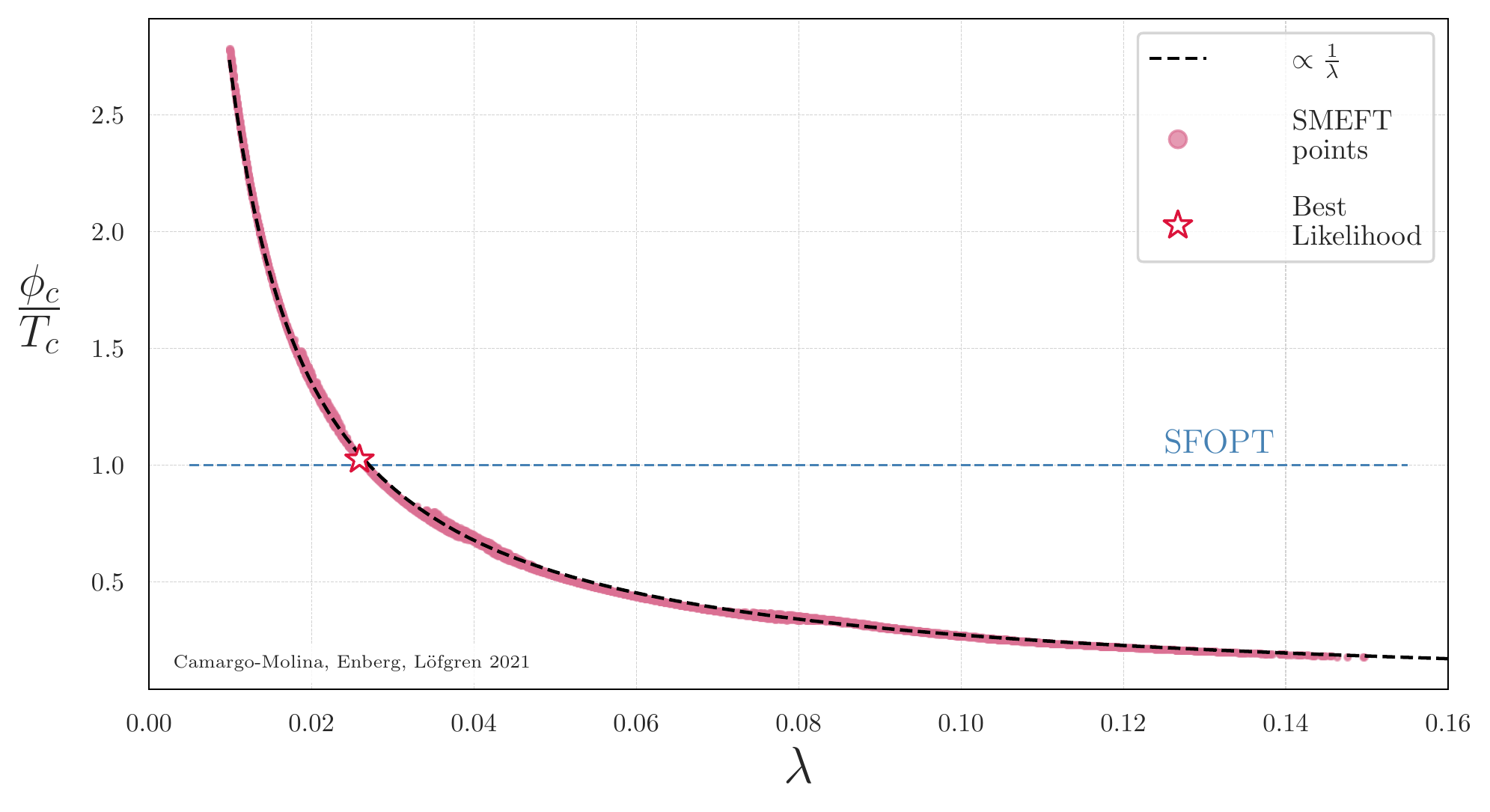} }}%
    \caption{$\phi_c/T_c$ as a function of $\lambda$ for the SMEFT parameter points in our study (shown as pink points). A fit proportional to $1/\lambda$ is shown with a black dashed line. $\phi_c/T_c > 1$ means a strongly first-order transition; the boundary is shown with a blue dashed line.}
    \label{Fig:PTstrength}%
\end{figure}

The results are shown in Fig.~\ref{Fig:PTstrength}, where we plot $\frac{\phi_c}{T_c}$ as a function of $\lambda$. A consistency check for the chosen parameter space is that the distribution of points goes roughly as $1/\lambda$, which is a direct consequence of \eqref{Eq:scaling} and noting that $ V^{}_{\textsc{LO}}(\phimin, T_c ) = 0$.

A few words have to be said about our decision to only consider the Higgs sector operators and impose a specific power counting when it comes to possible ultraviolet (UV) physics scenarios. There are heavy states that could give rise to the operators we consider here and no others when integrated out. For example, any pair of scalars that can contract into an $SU(2)$ doublet and color singlet, with total (sum) hypercharge equal to $1/2$, generate the three operators we consider after matching. With the appropriate hypercharge assignment they can also generate only those three operators and no others \cite{DasBakshi:2021xbl}. The admittedly necessary hierarchy between $\Cphi$ and $\CphiD$ (which has to be smaller than $10^{-7} \mbox{GeV}^{-2}$ due to electroweak precision tests) could simply arise from the details of the UV theory as $\Cphi$ can receive contributions from interactions that do not contribute to $\CphiD$ or $\Cphibox$. At the price of generating contributions to more operators, in some scenarios\footnote{For example, a quartic UV interaction with 3 Higgs fields and one heavy field carrying $(1, 2, \pm 3/2)$ $(\mathrm{SU}(3),\mathrm{SU}(2),\mathrm{U}(1))$ quantum numbers would contribute only to $\Cphi$, with a tree-level diagram, and not the other two operators} $\Cphi$ can receive a non-zero contribution already at the tree-level matching stage while $\CphiD \, \Cphibox$ only at one loop. A more contrived scenario, reminiscent of supersymmetry, would be the cancellation of contributions to $\CphiD$ from the two scalars mentioned above by contributions from two fermions with the same SM quantum numbers.\footnote{This works as $\CphiD$ would be the only operator receiving such fermionic loop contributions, see e.g. tables 5 and 6 of \cite{DasBakshi:2021xbl}}

Nevertheless, the main focus of this work is to explore the possibility, if at all, of having a first-order EWPT in the SMEFT with $\lambda > 0$ within the experimentally allowed parameter space. We will show in the following that the restriction to the Higgs sector SMEFT will be enough for that purpose. As such, we do not want to carry out an exhaustive analysis of the effects of many operators. Our results will then show the best-case scenario, as any departure from our minimal assumptions will result in either more constraints from other SMEFT operators or a failure to generate the required hierarchy that keeps $\CphiD$ small while keeping $\Cphi$ large enough to fit the Higgs mass when $\lambda$ is small enough to allow for a first-order PT. 

Thus a complete global fit of the 59-dimensional SMEFT parameter space is beyond the scope of this article, and we do not intend to carry out an exhaustive statistical analysis. A simple global fit including the most relevant experimental data using the sampling in \eqref{eq:sampling} will be more than sufficient to show that a first-order EWPT in the SMEFT is possible for $\lambda > 0$ in regions of the parameter space consistent with experimental data. 

\begin{figure}[tbh!]%
    {{\includegraphics[width = 0.7\textwidth]{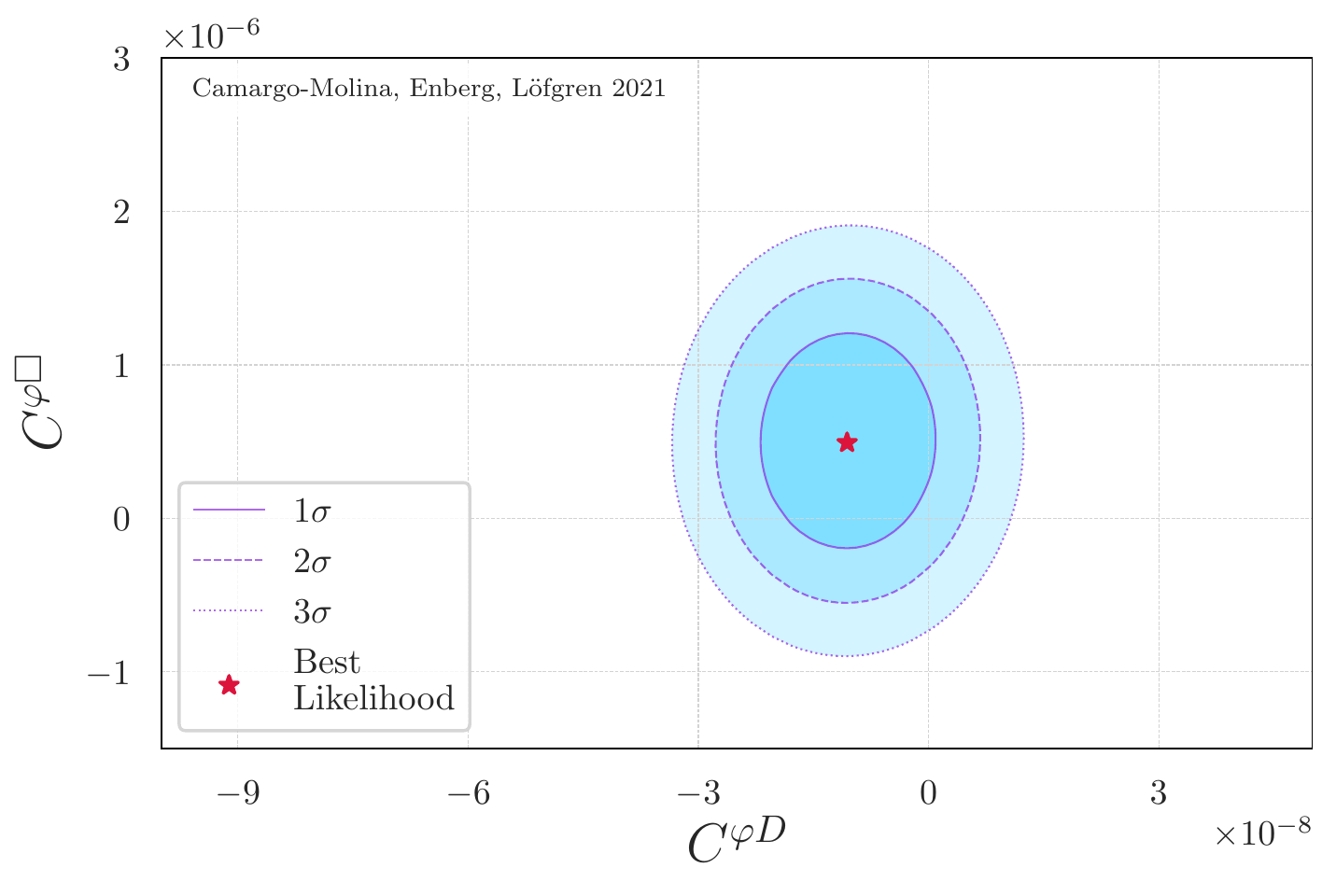} }}
    {{\includegraphics[width = 0.7\textwidth]{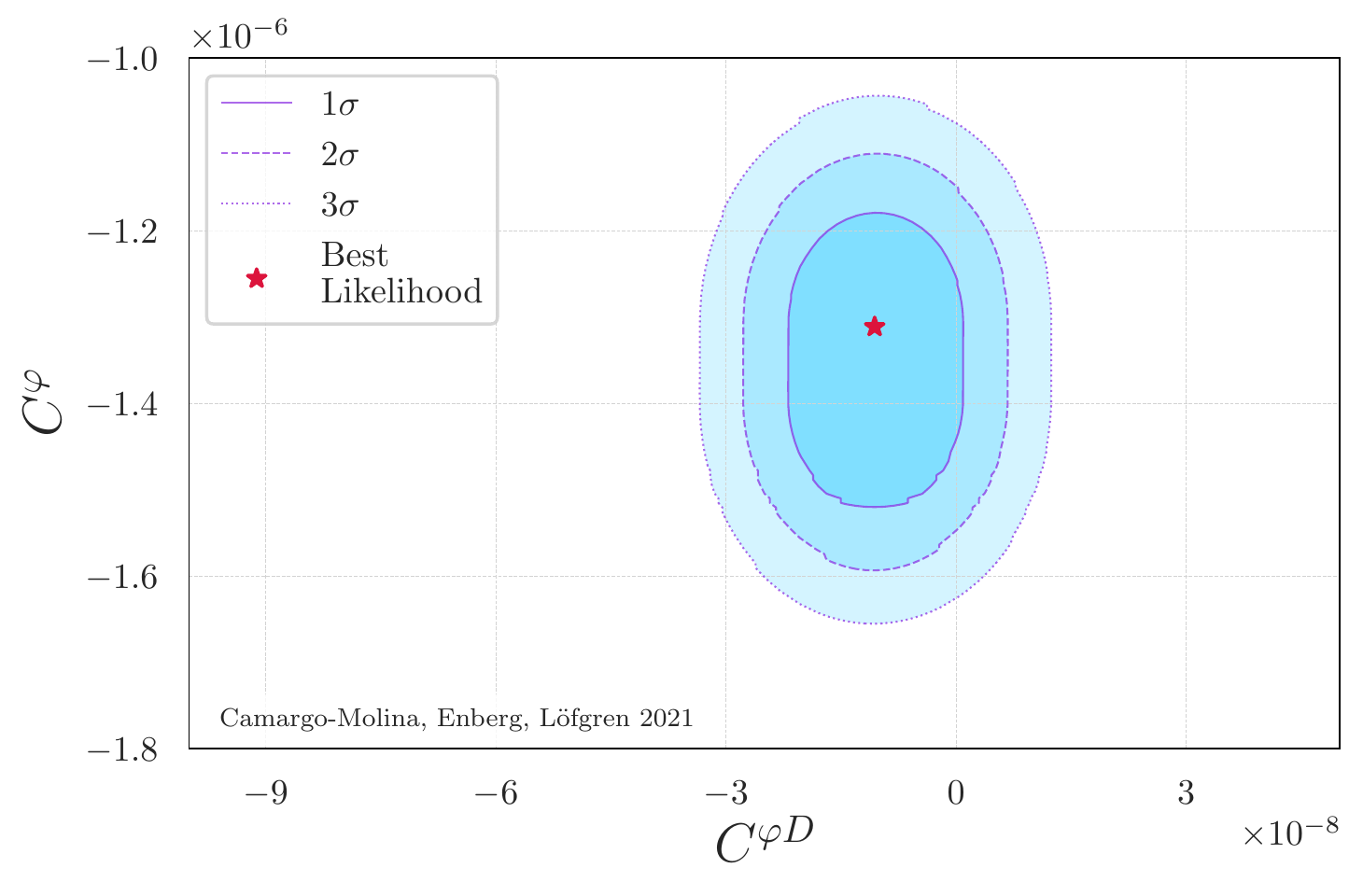} }}
    \caption{Profile likelihood for the $\CphiD$--$\,\Cphibox$ (top) and $\CphiD$--$\,\Cphi$ (bottom) planes for points leading to a strongly first order phase transition. One-, two- and three-$\sigma$ contours are drawn with full, dashed and dotted lines respectively. The point with the best likelihood and a strong FOPT is shown with a star.}
    \label{Fig:globalfit}%
\end{figure}

\begin{figure}[tbh!]%
    {\includegraphics[width = 0.9\textwidth]{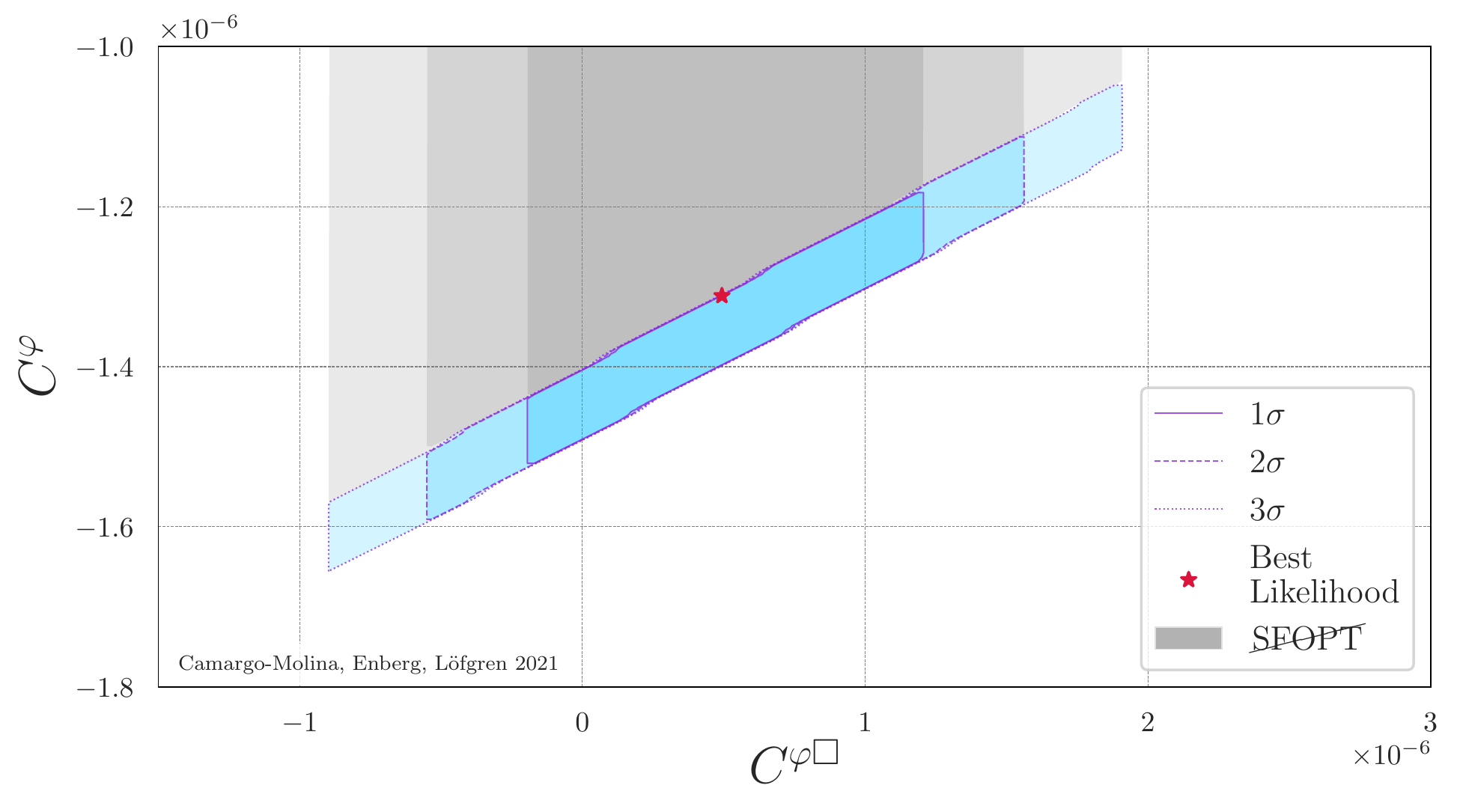} 
    \caption{Profile likelihood in the $\Cphibox$--$\,\Cphi$ plane. The intensity of the shading corresponds to the one-, two- and three-$\sigma$ regions. The blue-shaded region leads to a strong FOPT and the gray-shaded region does not. We only show contours for the region with a first order transition. The point with the best likelihood and a strong FOPT is shown with a star. The lower area is empty as points there have $\lambda<0$ and were not included in the study.}
    \label{Fig:globalfitBoxPhi}%
    }
\end{figure}

For this we use the software package \texttt{smelli} \cite{smelli}, a global likelihood calculator for the SMEFT which uses \texttt{flavio} \cite{flavio} for observable calculation and \texttt{wilson} \cite{wilson}\footnote{Note that we used a modified version. The changes included a bug fix in the Higgs mass calculation and an improved determination of $m^2$ and $\lambda$ using non-linearized (in the Wilson Coefficients) formulas for the Higgs mass and the field value at the minimum. This was implemented after discussion with the authors in the context of this work. The changes while not included in the latest release, have since been included in the latest commit in the GitHub repository \cite{wilsongithub}.} for RG running of Wilson coefficients and matching to the low-energy EFT. The \texttt{smelli} code includes a wide range of observables and experimental data, including EWPT and Higgs physics, see \cite{smelli} for a detailed list of observables. The results from our global fit are shown in Fig.~\ref{Fig:globalfit} and Fig.~\ref{Fig:globalfitBoxPhi}.

As can be seen in Fig.~\ref{Fig:globalfitBoxPhi}, the region with a strongly first order phase transition is clearly separated in the $\Cphi$--$\Cphibox$ plane, showing the importance of both coefficients in accommodating both the measured Higgs mass and a sufficiently small $\lambda$. 
As can be seen from Figure.~\ref{Fig:globalfit}, we find that at present $\Cphi$ is not significantly constrained by data. This is expected, as current constraints come from e.g.\ two-loop corrections to the W-boson mass \cite{Dawson:2020oco}. 
This is different for $\CphiD$ and $\Cphibox$, both of which are constrained mainly by Higgs production and electroweak precision tests. The bounds for $\CphiD$ are driven by the latter, which can be quickly understood from the correction to the $\rho$ parameter 
\begin{equation}
    \rho = \frac{M_W^2}{M_Z^2 \cos^2 \theta_W} =  1 - \frac{1}{2} \CphiD v^2 ,
\end{equation}
which to stay within $1 \sigma$ of the measured value already leads directly to a bound of  $-1.9 \times 10^{-8} \, \GeV^{-2} < \CphiD < -6.6 \times 10^{-9} \, \GeV^{-2}$. On the other hand the weaker bound on $\Cphibox$ is also driven by electroweak physics, such as $Z$-pole observables, but also by Higgs physics. For example, the coupling between the top quark and the Higgs is now 
\begin{equation}
\mathcal{L} \supset \bar{t} \left[ \frac{m_t}{\phimin} (1 - \frac{1}{4}\CphiD \phimin^2 + \Cphibox \phimin^2)\right] P_R \, t \, \phi
\end{equation}
which leads to modified cross-sections for processes such as $\bar{t}\, t \rightarrow h\rightarrow W^+ W^-$. Nevertheless, there is no problem in having $\Cphi$ doing the heavy lifting in raising the Higgs mass for small $\lambda$, and the allowed values are well within the range necessary for generating a wall at finite temperature and thus a first-order phase transition. 

\begin{figure}[tbh!]\label{fig:pulls}
    {\includegraphics[width = 0.7\textwidth]{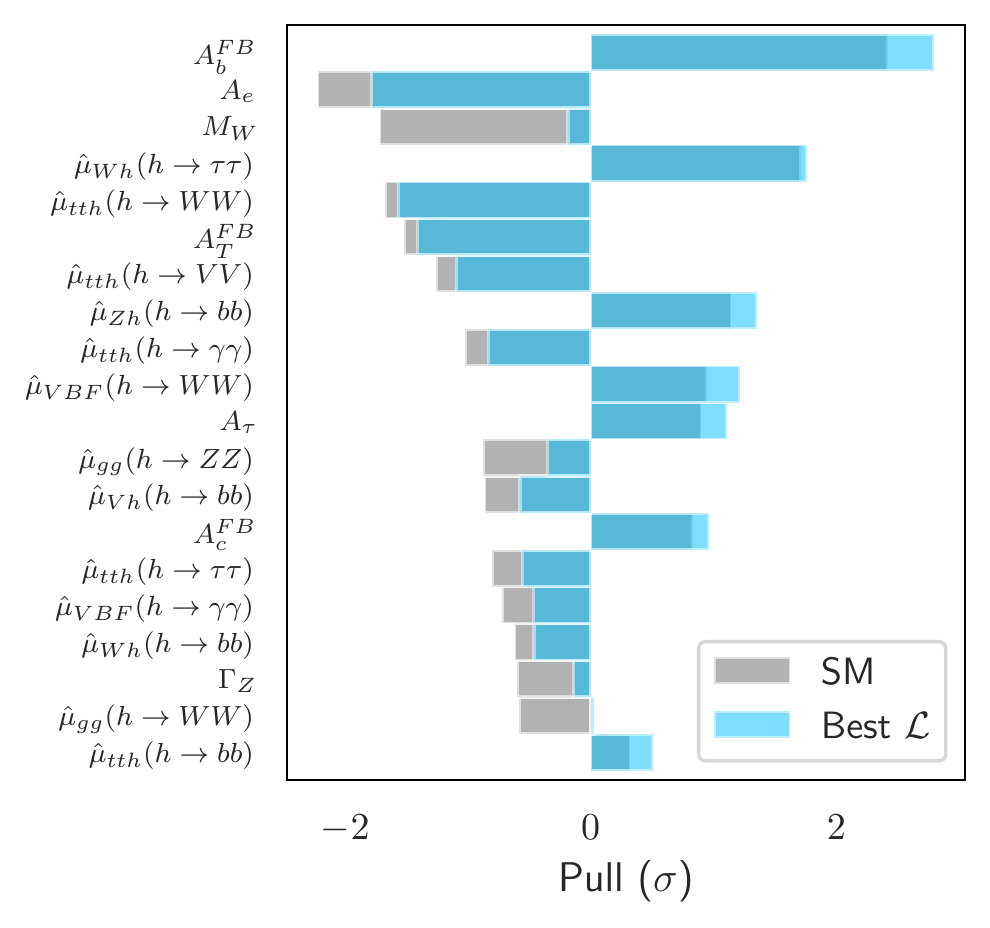} }
    \caption{ Pulls for individual observables for the point with best likelihood in our scan, within the region with a strong FOPT. We show the twenty observables deviating the most from the SM prediction.}
\end{figure}

We note that the point in the scan that has the best value of the likelihood and a strong FOPT fits data modestly better than the SM ($\sim1\sigma$). In Fig.~\ref{fig:pulls} we show the pulls for the 20 observables where the SMEFT prediction deviates the most from its SM counterpart. This is meant as an illustration of the phenomenology at hand in this region of the parameter space. We note the overall better fit of several electroweak precision observables, in particular the $W$  mass, which is usually a stringent bound on other SMEFT scenarios. 
 
\section{Implications for di-Higgs production at colliders}\label{sec:dihiggs}

\noindent{}One of the goals of the High Luminosity LHC (HL-LHC) experiment is to probe the Higgs self couplings. A promising avenue to set limits on such interactions is Higgs pair production, or di-Higgs production, where the LHC at its current luminosity has little sensitivity. The current LHC will not observe di-Higgs production if the SM prediction is correct, but some new physics models predict significantly larger cross sections. 

As we have seen in the preceding sections the existence of a first-order EWPT in the SMEFT depends sensitively on both $\lambda$ and $\Cphi$, where the latter parametrizes new physics effects on Higgs couplings. Both $\lambda$ and $\Cphi$  also set the value of the triple Higgs coupling. It is therefore interesting to look at the connection between a first-order EWPT and di-Higgs production in SMEFT.

Consider the triple Higgs self-coupling. In the SMEFT the trilinear interaction of the Higgs has a momentum-dependent part coming from the operators $\QphiD$ and $\Qphibox$. Let us here for simplicity ignore that part, since the corresponding Wilson coefficients are smaller. Then we can consider the momentum-independent part of the interaction and define a SMEFT triple Higgs coupling $\lambda_{HHH}$ compared to the SM triple Higgs coupling $\lambda_{HHH}^\text{SM}$. Let us express the ratio in terms of the Higgs mass instead of $\lambda$. If we neglect $\CphiD$ and $\Cphibox$, we get 
\begin{equation}
    \frac{\lambda_{HHH}}{\lambda_{HHH}^\text{SM}} = 1 - 2 \frac{\Cphi v^4}{m_h^2}.
\end{equation}
At the LHC, the recently released best limit is $-1.5<\lambda_{HHH}/\lambda_{HHH}^\text{SM}<6.7$ \cite{ATLAS-CONF-2021-016}. Moreover, in \cite{Kim:2018uty} an optimistic case is made for limits on $\lambda_{HHH}$ at the HL-LHC, leading to the bounds $0.2 < \lambda_{HHH} / \lambda_{HHH}^\text{SM} < 2.3$ at 68\% C.I.
Using these values we can then find a rough estimate of the bound on $\Cphi$ from the LHC and the possible bound from optimistic prospect at the HL-LHC:
\begin{align}
-1.2 \cdot 10^{-5} &< \Cphi_{\text{\tiny LHC}} < 5.3 \cdot 10^{-6}\\ \nonumber
-2.8 \cdot 10^{-6} &< \Cphi_{\text{\tiny HL-LHC}} < 1.7 \cdot 10^{-6}.
\end{align}

Setting limits on the Higgs triple coupling is famously difficult, but the naive estimation above shows that the limits are precisely in the region of interest for a strong FOPT with positive $\lambda$. 

\begin{figure}\label{Fig:FeynmanLHC}
 \centering
  \begin{minipage}{\linewidth}
      \raisebox{50pt}{a)}{\includegraphics[width=0.3\linewidth]{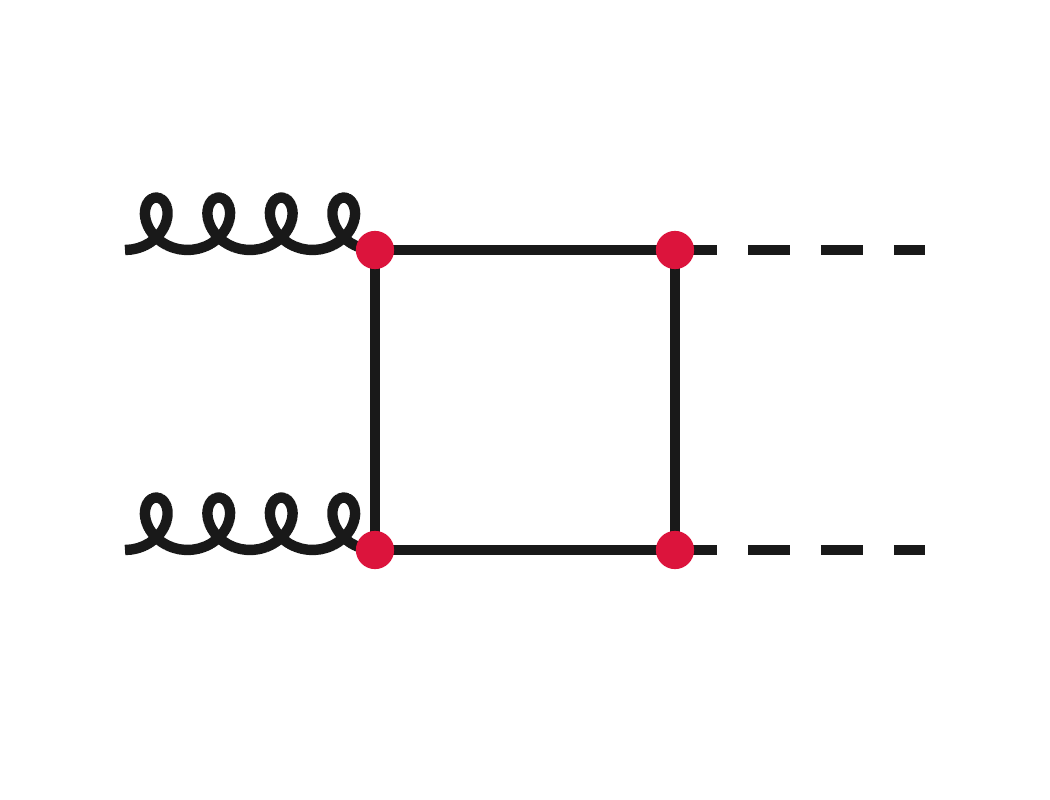}}\hfill
      \raisebox{50pt}{b)}{\includegraphics[width=0.3\linewidth]{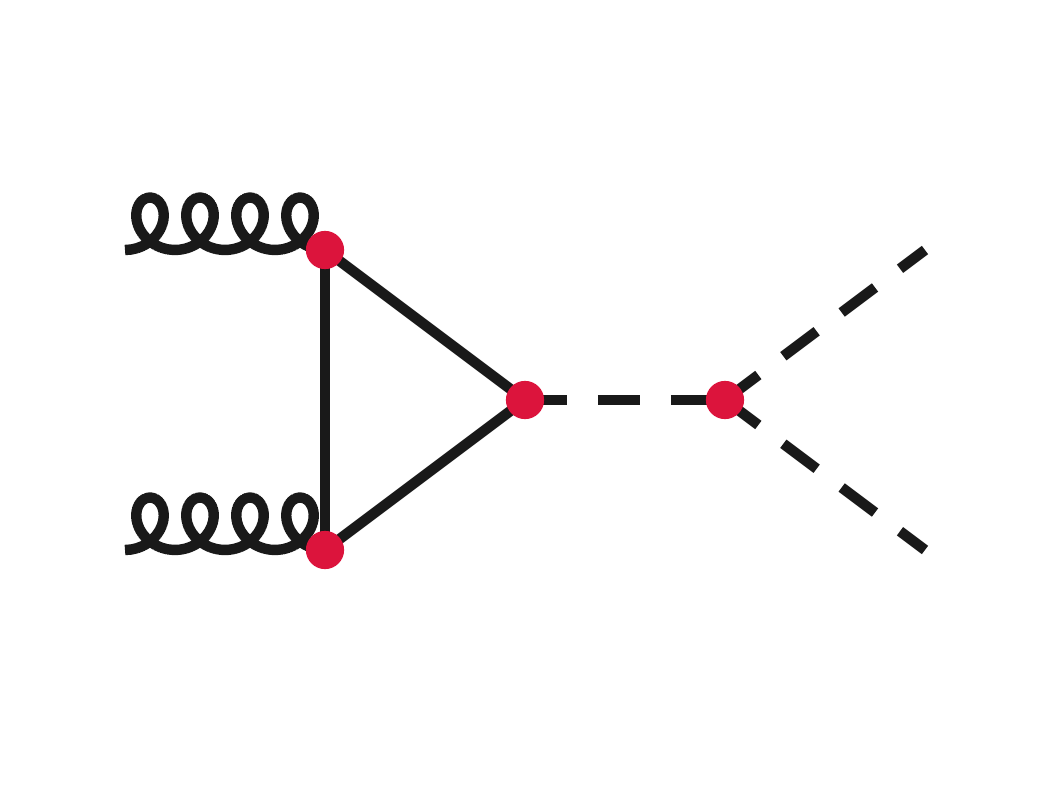}}\hfill
      \raisebox{50pt}{c)} {\includegraphics[width=0.3\linewidth]{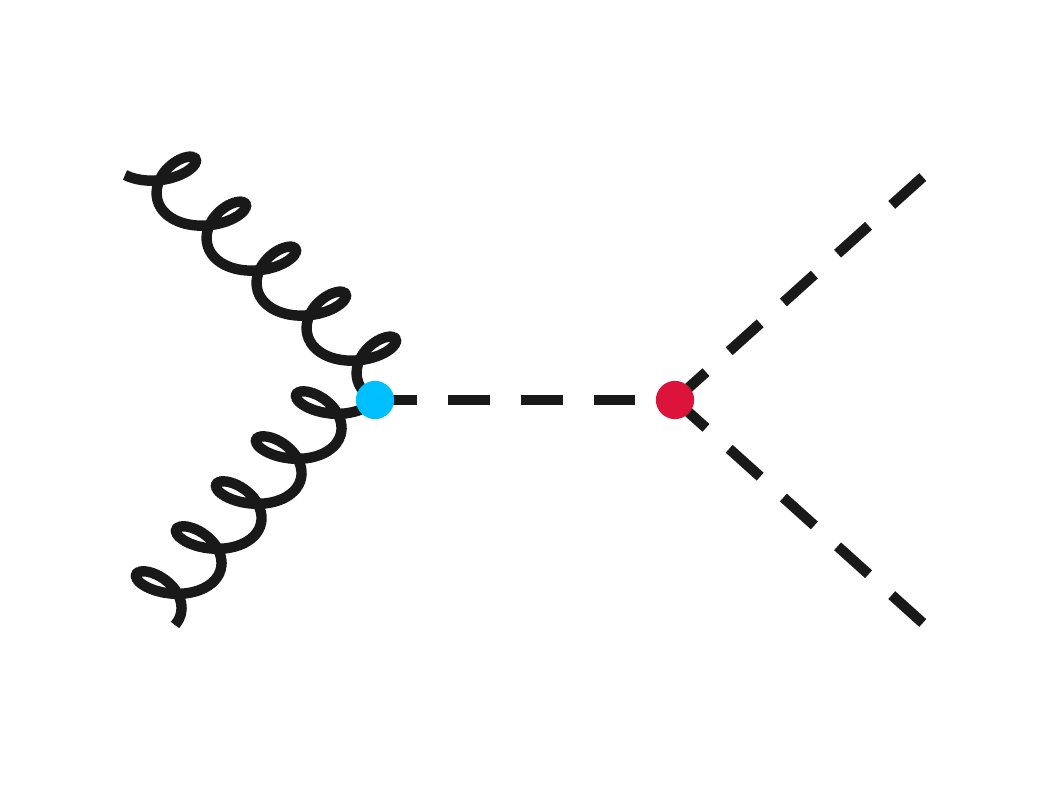}}\\
      \raisebox{50pt}{d)}{\includegraphics[width=0.3\linewidth]{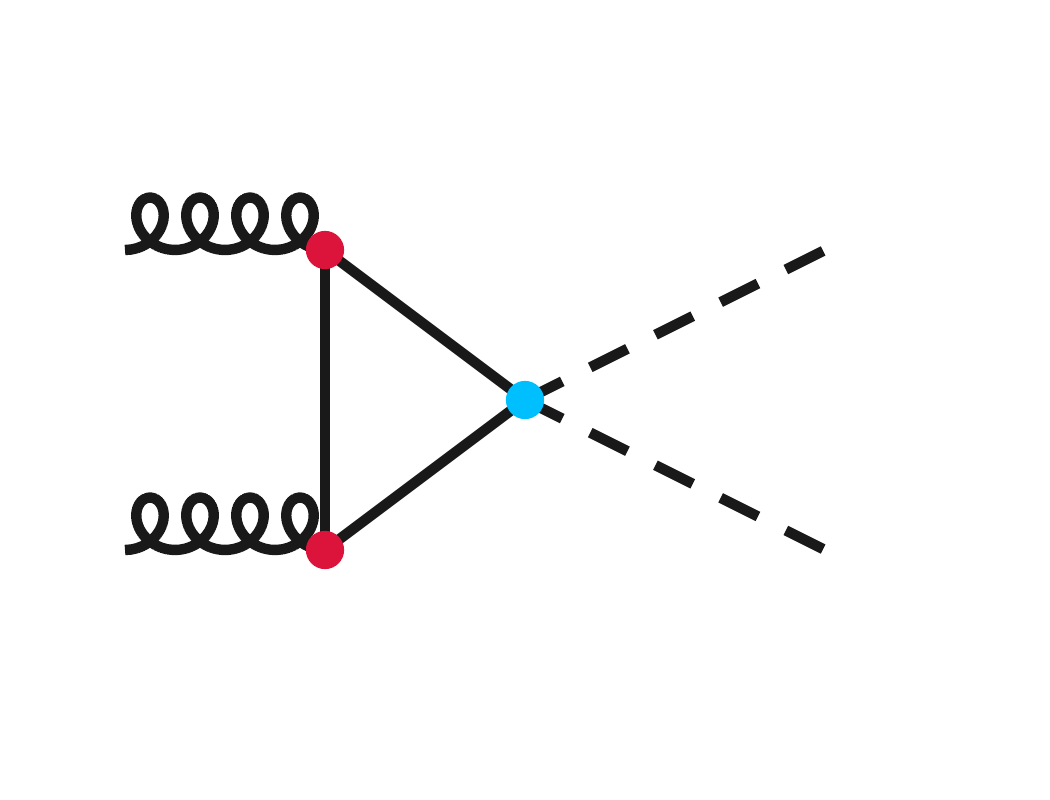}}\hfill
     \raisebox{50pt}{e)}{\includegraphics[width=0.3\linewidth]{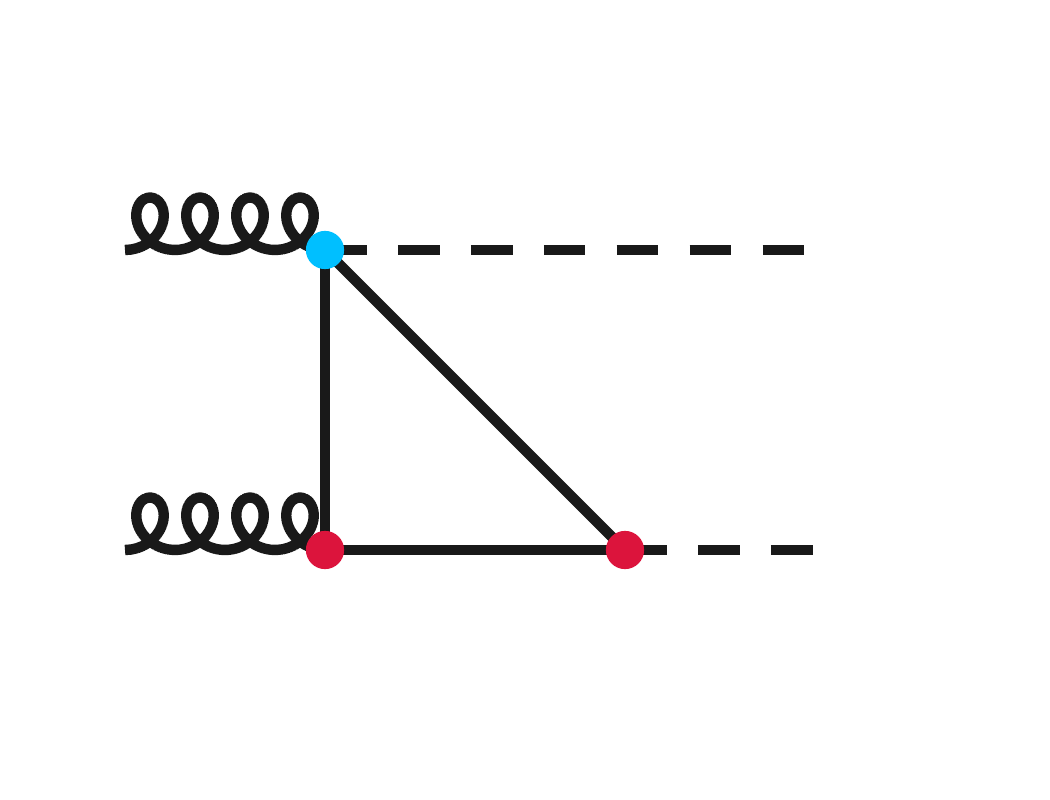}}\hfill
     \raisebox{50pt}{f)}{\includegraphics[width=0.3\linewidth]{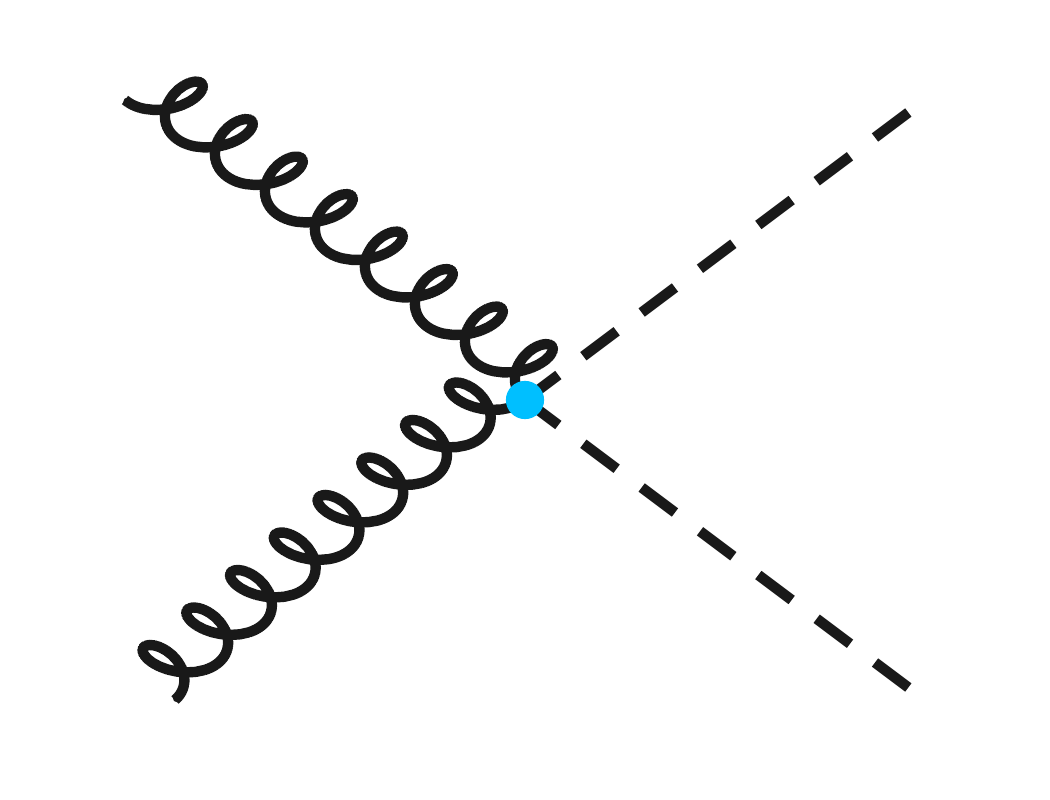}}
  \end{minipage}
\caption{Feynman diagrams relevant for Di-Higgs production at the LHC. A red vertex indicates interactions present in the SM but modified by the SMEFT operators. A blue vertex indicates interactions not present in the SM.}
\end{figure}

The limits above were obtained by studying double-Higgs production at colliders, where the triple Higgs coupling plays an important role. We want to note that our discussion here is somewhat superficial. A thorough study of di-Higgs production at the LHC and HL-LHC would require the consideration of the full set of operators. See e.g.\ Ref.~\cite{Corbett:2017ieo} for such a discussion. There are many relevant processes that play a role, and we show the most relevant ones in Figure~\ref{Fig:FeynmanLHC}. Nevertheless, we think there is a strong case for more sophisticated studies focusing on this, as the current bounds lie very closely to either ruling out or supporting a strong FOPT due to heavy physics. 
In the medium-long term, at future linear colliders, prospects are even more promising, as the cleaner signals simplify the probing of deviations in the Higgs self-couplings. If future bounds are consistent with a strong FOPT, stringent bounds will help in weeding out unfavorable high-energy scenarios and shed light on the details of baryogenesis. Moreover, a better understanding of the allowed SMEFT parameter space will also help in searching for gravitational waves signals coming from the EWPT, as the frequency and spectrum strongly depend on the same Wilson coefficients.

\section{Conclusions}\label{sec:conclusions}
 
\noindent{}In this work we have shown the importance of studying the region of the SMEFT parameter space that leads to a first order electroweak transition by means of radiative corrections.
This region is characterized by a very small but positive Higgs quartic coupling $\lambda$. This region allows for a first-order transition for Wilson coefficients consistent with a larger cutoff scale and in turn a more consistent use of EFTs compared to the negative $\lambda$ case that has been often studied previously. We are optimistically suggesting that a larger family of BSM theories that describes a first-order transition should be tested using the SMEFT framework than was known before. We have also briefly touched on the possibilities for new physics scenarios that would generate the Wilson coefficients at the center of our study. We have also shown that a large portion of the parameter space is in agreement with experimental data. Our best fit point (within the region with a first-order EWPT), while having sizable Wilson Coefficients, also fits the data slightly better than the SM. 

We should not forget that the SMEFT is a parametrization of heavy new physics, best thought of as a proxy for studying a wide family of BSM theories. We find that this new region of parameter space brings forward a strong case for testing whether a first-order phase transition might be responsible for the matter-antimatter asymmetry in the near future. Our results also show that such a test, which translates to setting limits on $\Cphi,\Cphibox$ and $\CphiD$, can be tackled on two fronts. 

First, in collider physics, where probing the Higgs boson self-interactions is at the forefront of current and future efforts. We have shown that the SMEFT parameter space within reach of such studies includes the boundary between a strong and weak first order electroweak phase transition. That is true already for the HL-LHC and perhaps could be achieved at the LHC with an improved analysis. 

Second, using gravitational waves:\ a measurement of the stochastic gravitational wave background consistent with a first-order transition can be translated into limits on the SMEFT Wilson coefficients. The prospects for such a measurement are very promising with the planned launch of LISA in 2031. While it is true that a larger new physics scale might in principle lead to fainter GW signals, there is still a lively discussion in the literature regarding the precision and accuracy of current predictions for spectra visible at LISA~\cite{Croon:2020cgk,Gould:2021oba}. 

Most importantly, this illustrates the relevance of the interplay between particle physics and cosmology. The physics of the EWPT links both fields experimentally and our work supports the idea of using collider physics and gravitational waves to test the same early-universe phenomena. The long-term future of high energy physics depends on the prospects for getting definitive answers to fundamental questions. Explaining the baryon asymmetry might very well be the most compelling possibility for such an answer. In this work we show that the first step, namely a determination of the order of the EWPT, is within reach of the LHC and future colliders.

For future work, there are several issues remaining to be addressed. First, it has recently been argued that more care should be applied when studying the electroweak phase transition in order to reduce theoretical uncertainties like renormalization-group scale dependence~\cite{Croon:2020cgk}. A future refinement of our study should use the methods of dimensional reduction which can automatically deal with resummations and can even be combined with lattice data to capture non-perturbative effects (for recent pedagogical examples of dimensional reduction, see e.g.~\cite{Croon:2020cgk,Gould:2021dzl,Schicho:2021gca,Niemi:2021qvp}). It should also study quantities related to the nucleation of bubbles more quantitatively. Second, it would be interesting to match the Wilson coefficients to actual UV theories, to get an understanding of whether the parameter space we have studied can arise in sensible extensions of the SM---just like in e.g.\ the recent study in Ref.~\cite{Postma:2020toi}. Finally, the prediction for di-Higgs production in the SMEFT should be explored in this region of parameter space.

Based on our results in this paper, we might speculate that if it will turn out that the EWPT is indeed first-order, then either there should be some new bosonic states around the electroweak scale that modify the effective potential, or the new physics that allows the EWPT to take place is heavier, and in this case, we would expect the quartic Higgs coupling to be small but positive.

\acknowledgments

\noindent{}This work was funded in part by the Carl Trygger Foundation through grant no.\ CTS 17:139 and in part by the Knut and Alice Wallenberg foundation under grant no.\ KAW 2017.0100. The computations were enabled by resources in projects SNIC 2019/8-224 and SNIC 2021/22-106 provided by the Swedish National Infrastructure for Computing (SNIC) at UPPMAX, partially funded by the Swedish Research Council through grant agreement no.\ 2018-05973. We thank Peter Stangl for helpful communications regarding \texttt{smelli}, \texttt{wilson} and the determination of the Higgs potential parameters. We thank Oliver Gould and Graham White for their comments on the draft.

\bibliographystyle{JHEP}
\bibliography{bib.bib}

\end{document}